\documentclass[12pt,preprint]{aastex}
\usepackage[numberedappendix]{emulateapj5}  

\slugcomment{accepted for publication in ApJ} 

\newcommand\rss{{\rm \scriptscriptstyle}}
\newcommand\etal{{et al.}\thinspace}
\newcommand\be{\begin{equation}}
\newcommand\ee{\end{equation}}
\newcommand\bea{\begin{eqnarray}}
\newcommand\eea{\end{eqnarray}}

\newcommand\lLr{{$ \langle L \rangle$}\thinspace}

\shorttitle{AGN heating and SZ effect}
\shortauthors{Roychowdhury et al.}
 
\begin{document}
 
\title{AGN heating, thermal conduction and Sunyaev-Zeldovich effect in galaxy
groups and clusters}

\author{S. Roychowdhury}
\affil{Raman Research Institute, Bangalore 560080, India;
\email{suparna@rri.res.in}}
\author{M. Ruszkowski \altaffilmark{1} and B.B. Nath\altaffilmark{2}}
\affil{JILA, Campus Box 440, University of Colorado at Boulder, CO
80309-0440; \email{mr,biman@quixote.colorado.edu}}               
\altaffiltext{1}{{\it Chandra} Fellow}
\altaffiltext{2}{JILA Visiting Fellow, on leave from Raman Research Institute}          

\begin{abstract}
We investigate in detail the role of active galactic nuclei (AGN) on the
physical state of the gas in galaxy groups and clusters, and the implications
for anisotropy in the cosmic microwave background (CMB) from Sunyaev-Zeldovich
(SZ) effect. We have recently showed that AGNs can significantly change the
entropy of the intracluster medium (ICM) and explain the observations of 
excess entropy in groups and clusters. AGNs are assumed to deposit energy via
buoyant bubbles which expand as they rise in the cluster atmosphere and do
$PdV$ work on the ICM. Here, we include the effect of thermal conduction, and
find that the resulting profiles of temperature and entropy are consistent
with observations. 
Unlike previously proposed models, our model predicts that isentropic
cores are not an inevitable consequence of preheating. The model also 
reproduces the observational trend for the density profiles to flatten 
in lower mass systems.

\indent
We deduce the energy $E_{\rm agn}$ required to explain the entropy
observations as a function of mass of groups and clusters $M_{\rm cluster}$ 
and show that $E_{\rm agn} \propto M_{\rm cluster}^{\alpha}$ with $\alpha\sim 1.5$.
We demonstrate that the entropy measurements, in conjunction with our model,
can be translated into constraints on the cluster---black hole mass relation. 
The inferred relation is nonlinear and has
the form $M_{\rm bh}\propto M_{\rm cluster}^{\alpha}$. This scaling
is an analog and extension of a similar relation between the black hole mass and 
the galactic halo mass that holds on smaller scales. 

\indent
In addition, we
study the implications of these results for thermal SZ effect. We show that
the central decrement of the CMB temperature is reduced due to the enhanced
entropy of the ICM, and that the decrement predicted from the plausible range 
of energy input from the AGN is consistent with available data of SZ decrement.
We also estimate the Poisson contribution to the angular 
power spectrum of the CMB from the SZ effect due to AGN heating.
We show that AGN heating, combined with the observational constraints on 
entropy, leads to suppression of higher multipole moments in the power spectrum
and we find that this effect is stronger than previously thought. The supression
in the power spectrum in our model is due to depletion of gas from the central
regions that is more efficient in low mass clusters and groups than in massive 
clusters.

\end{abstract}


\section{Introduction}

The formation of structures in the Universe is believed to be hierarchical, 
as primordial density fluctuations, amplified by gravity, collapse and merge 
to form progressively larger systems. This hierarchical development leads to 
the prediction of self-similar scaling relations between systems of different masses 
and at different epochs (Peebles, 1980). These structures contain two components 
$\hbox{--}$ the gravitationally dominant dark matter and the baryons contained in 
these potential wells whose response to processes other than gravitational interactions 
bring about deviations from the self-similar scaling relations.

Clusters and groups of galaxies contain dark matter and hot, diffuse gas called 
the intracluster medium. It was believed that this intracluster gas 
follows a self-similar scaling relations. 
However, recent observations of clusters and groups of galaxies have shown 
that the scaling relations are not self-similar. The observed relations of different physical 
parameters of the ICM such as density, temperature, X-ray luminosity and entropy have 
mostly confirmed the requirement for non-gravitational processes like AGN heating 
and radiative cooling (Lloyd-Davies et al 2000, Ponman et al. 2003, 
Sanderson et al. 2003, Pratt \& Arnaud 2003, Pratt \& Arnaud 2005). Simulations and 
theoretical models of clusters with gravitational processes alone also point to the 
fact that the entropy or X-ray luminosity observations can be matched only with 
non-gravitational heating (Roychowdhury \& Nath 2003). Many theoretical models have 
been proposed to explain these X-ray observations by heating from supernovae (Valageas \& 
Silk 1999; Wu, Fabian \& Nulsen 2000), radiative cooling (Bryan 2000; Voit \& Bryan 2001; 
Muanwong et al. 2002; Wu \& Xue 2002a; Dav\'e, Katz \& Weinberg 2002, Tornatore et al. 2003), 
accretion shocks (Tozzi \& Norman 2001; Babul et al. 2002), quasar outflows (Nath \& 
Roychowdhury 2002), and ``effervescent heating'' (Roychowdhury et al. 2004; hereafter 
RRNB04). More information on various heating models
can be found in a review by Gardini \& Ricker (2004).

In a recent study Croston et al. (2004) presented the luminosity-temperature 
relation for groups and separated the sample into ``radio loud'' or ``quiet'' objects. 
They showed that ``radio loud'' groups deviated more from the self-similar scaling relation than the 
radio quiet ones. 
\footnote{The extrapolation of the cluster L-T relation onto Figure 2 of 
Croston et al. (2004) would fall above all the points. Slightly smaller deviation
in the radio quiet AGN could be due to the fact that they are not active now but were 
active in the past. (We thank the referee for pointing out this possibility, 
see also Donahue (2005))}
This argument adds more credibility to the 
idea that AGNs are responsible for the entropy ``floor'' and deviations from self-similar 
scaling relations.

In a recent paper, we have studied the effect of buoyant bubbles which
deposit energy into the ICM as they rise and expand in the cluster atmosphere
(RRNB04). We studied the evolution of the density, temperature and entropy
profiles of the ICM under the effect of ``effervescent heating'' from buoyant
bubbles, radiative cooling and convection, and deduced the energy required
from the AGN to satisfy the entropy observations of the ICM. Although the
constraints from entropy observations at two different radii ($0.1 r_{200}$
and $r_{500}$) were satisfied, the temperature and entropy profiles were
somewhat different from observed profiles, in that the entropy profiles had
a flattened core and the temperature in the inner regions were enhanced
(see RRNB04 for details). To resolve this issue, we study the effect of
thermal conduction  in this paper.

The role of thermal conduction in the intracluster medium has  
generated a lot of interest in recent times. Its role, however,
has mostly been discussed or studied in detail in reference to cooling flows 
in the central regions of clusters. Many authors have investigated whether 
thermal 
conduction alone can act as a heating source in the centre of clusters to stop 
the gas from radiatively cooling to very low temperatures (e.g., 
Zakamska \& Narayan, 2003; 
Voigt \etal 2002; Loeb 2002). It has also been studied along with other 
heating mechanisms like ``effervescent heating'' again in the context of cooling 
flows in the centers of clusters (Ruszkowski \& Begelman, 2002). 
In this paper, we study the effect of thermal conduction along with heating
via AGN in changing the entropy profile of the ICM.

Until recently it was only X-ray observations that yielded information 
about the entropy excess. Due to advances in detectors and new observing strategies 
(Birkinshaw 1999, Grego et al. 2001; Grainge et al. 2002; Reese et al. 2002; 
Zhang \& Wu 2000), the thermal Sunyaev-Zeldovich (SZ) effect (Sunyaev \& 
Zeldovich 1972, 1980) is emerging as an  {\it independent} test of the density 
and the thermal structure of clusters, thus equivalently of the entropy excess. 

Many authors have investigated the role of excess entropy in clusters on the 
SZ effect and tried to quantify it (White et al. 2002; Springel et al. 2001; 
da Silva et al. 2001, 2004; Cavaliere \& Menci 2001; Holder \& Carlstrom 2001 
\& McCarthy et al. 2003a, 2003b). Holder \& Carlstrom (2001), Cavaliere 
\& Menci (2001) and McCarthy et al. (2003a) have also examined a few SZ scaling 
relations for individual clusters. They have shown that the SZ decrement 
is reduced in individual clusters as a result of energy injection and that the 
SZ scaling relations deviate from the self-similar predictions. In a more 
recent effort, Lapi et al. 2003 have estimated the enhancements in the SZ effect 
due to transient blastwaves from quasars and the depressions when the hydrostatic 
equilibrium is recovered. 

In this paper we further explore the consequences of heating the intracluster medium 
via the ``effervescent heating'' mechanism (Begelman 2001, Ruszkowski 
\& Begelman 2002, RRNB04) and thermal conduction. We also focus on 
the SZ decrement resulting from this heating model and 
calculate the angular power spectrum of the CMB due to  
effervescent heating and thermal conduction. Using the entropy data, 
we also put constraints on the 
relation between the black hole mass and the cluster mass. We show that 
it is has to be nonlinear in analogy to the black hole---galaxy halo relation
that is expected to hold on smaller scales.

The paper is organized as follows. In Section 2 we briefly describe our model.
In Section 3 we estimate the central SZ decrements. 
We also simulate the evolution of the central SZ signal due to AGN
heating, cooling and conduction. In Section 4 we estimate the angular power spectrum 
of the SZ temperature decrement in our models. We present our results and discussion 
in Section 5. In Section 6 we discuss the relation between the mass of the
central black hole and the mass of the cluster.
Finally, our conclusions are summarized in Section 7.    

We assume throughout the paper $\Omega_{\rm\Lambda} \, = \, 0.71$, $\Omega_{0} \,
=\,0.29$, $\Omega_{\rm b} \,={\,0.047}$ and $h\,=\,0.71$ which are the best fit 
parameters from WMAP (Spergel et al. 2003).

\section{Model of the intracluster medium}

\subsection{The default state of the ICM}

The details of the initial conditions of our model are similar to those adopted 
by RRNB04. In brief, we assume that the ICM is characterized by a 
``universal temperature profile'' in the range $0.04\le r/r_{\rm vir}\le 1$ 
given by 

\begin{equation}
\frac{T_{\rm o}}{\langle T\rangle}=\frac{b}{(1+r/a)^{\delta}}
\end{equation}

\noindent
where $\langle T\rangle$ is the emission-weighted temperature, $b=1.33$, $a=r_{\rm vir}$
and $\delta = 1.6$.

This choice of the
temperature profile as the initial temperature profile stems from the
results of recent simulation of adiabatic evolution of clusters (Loken et al. 2002).
Since our goal is determine the effect of energy input from AGN, the most natural
choice for an initial profile is that which arises from cluster evolution
without energy input. Also, as Roychowdhury \& Nath (2003) have shown, this
initial temperature profile results in entropy profiles closer to observations
than other profiles used previously in literature. The choice of this 
temperature profile therefore decreases the discrepancy in entropy and
therefore provides more conservative estimate of the energy required from AGN.

\indent
Density profiles are computed
assuming hydrostatic equilibrium of the gas in the background dark matter potential.
The background dark matter density profile is given by the self-similar Navarro, 
Frenk \& White (NFW) profile (Komatsu \& Seljak 2001) with a softened core 
(Zakamska \& Narayan, 2003)

\begin{equation}
\rho_{\rm dm} = \frac{ \rho_{\rm s} }{ {(r+r_{\rm c})(r+r_{\rm s})^2} },
\end{equation}

\noindent
where $r_{\rm s}$ is the standard characteristic radius of the NFW profile, 
$r_{\rm c}$ is a core radius inside which the density profile is a constant and 
$\rho_{\rm s}$ is the standard characteristic density of the usual NFW profile.
The mass profile is given by

\begin{equation}
M_{\rm dm}(\le r)\,=\, 4 \pi \rho_{\rm s}r_{\rm s}^3m(x),
\end{equation}

\noindent
where $m(x)$ is a non-dimensional mass profile

\begin{eqnarray}
m(x) & = & \frac{x_{\rm c}^2}{(1-x_{\rm c})^2}\ln (1+x/x_{\rm c}) \nonumber \\
     & + & \frac{(1-2x_{\rm c})}{(1-x_{\rm c})^2} \ln (1+x)-
           \frac{1}{1-x_{\rm c}}\frac{x}{1+x},
\end{eqnarray}

\noindent
where $x=r/r_{\rm s}$ and $x_{\rm c}=r/r_{\rm c}$. If $r_{\rm c}=0$, the usual 
mass distribution is recovered as in Komatsu \& Seljak (2001).
We follow Zakamska \& Narayan (2003) and assume $r_{\rm c}=r_{\rm s}/20$. This is  
a reasonable choice as cluster lensing studies suggest that the core radius can be 
$\sim$ tens of kilo-parsecs (Tyson et al., 1998; Shapiro \& Iliev, 2000).
We investigate the effect of the smoothing of the dark matter profile 
on our results.   

\indent
We do not model the effects of magnetic fields explicitly. However,
magnetic fields are likely to be below the equipartition value and thereby
dynamically unimportant (Fabian et al. 2002). 
Polarization measurements suggest that the magnetic fields in the vicinity of bubbles 
have subequipartition values (Blanton et al. 2003).
Nevertheless, they may still be important for suppressing instabilities 
on bubble-ICM interfaces. Such effects may implicitely be
included in the bubble heating model.

\subsection{Heating, cooling and thermal conduction}

\subsubsection{Effervescent heating and radiative cooling}

The effervescent heating mechanism is a gentle heating mechanism
in which the cluster gas is heated by buoyant bubbles of 
relativistic plasma produced by central AGN (Begelman 2000, 
Ruszkowski \& Begelman 2002). The average volume heating rate is 
a function of the ICM pressure gradient and is given by

\begin{equation}
{\cal H}=-h(r)P_{\rm gas}^{(\gamma_{\rm b}-1)/\gamma_{\rm b}}
\frac{1}{r}\frac{d\ln P_{\rm gas}}{d\ln r},
\end{equation}

\noindent
where $P_{\rm gas}$ is the ICM pressure, $\gamma_{\rm b}$ is the adiabatic index 
of buoyant gas in the bubbles and $h(r)$ is the normalization function 

\begin{equation}
h(r) = \frac{\langle L \rangle}{4 \pi r^2}[1-e^{-r/r_{\rm 0}}]q^{-1}.
\label{eq:normalisation}
\end{equation}

\noindent
In equation~(\ref{eq:normalisation}), $\langle L \rangle$ 
is the time averaged energy injection 
rate and $r_{\rm 0}$ is the inner heating cut-off radius. The 
normalization factor $q$ is defined by

\begin{equation}
q = \int_{r_{\rm min}}^{r_{\rm max}}P^{(\gamma_{\rm b}-1)/\gamma_{\rm b}}
    \frac{1}{r}\frac{d\ln P}{d\ln r}(1-e^{-r/r_{\rm 0}})dr ,
\end{equation}

\noindent
where $r_{\rm{max}}=r_{200}$.
The inner heating cut-off radius is the transition region between the 
bubble formation region and the buoyant (effervescent) phase. It can be determined 
self-consistenly by taking into account energy losses due to bubble creation,

\begin{equation}
\int_{\rm 0}^{r_{0}} 4\pi r^2 {\cal H} dr = {(\gamma_{\rm b}-1)
\over {\gamma_{\rm b}}} \langle L \rangle 
= \frac{1}{4}\langle L \rangle
\end{equation}

\noindent
This comes from the fact that the energy lost within the radius $r_{\rm 0}$ due 
to bubble creation is $P_{\rm 0}V_{\rm 0}$ where 
$V_{\rm 0}$ is the volume of the bubble at the creation region or 
transition region, $r_{\rm 0}$, and $P_{\rm 0}$ is the pressure at that region. 
However, the energy lost due to bubble expansion, i.e. in the effervescent phase is

\begin{equation}
\int_{P_{\rm 0}}^{0} P_{\rm bubble} dV .
\label{eq:effervescent_loss}
\end{equation}

\noindent
Assuming adiabatic evolution of the gas inside the 
bubbles (low density, low radiative losses) and mass conservation in the bubble one 
has $dV=(1/\gamma_{\rm b})(P_{\rm 0}/P_{\rm bubble})^{1/\gamma_{\rm b}} 
V_{\rm 0}dP/P$. On integrating, one gets effervescent energy as 
$3P_{\rm 0}V_{\rm 0}$ for $\gamma_{\rm b}=4/3$. It can be easily seen from 
above that the energy loss due to bubble creation is approximately 25$\%$ of 
the total energy available for heating. Thus this condition sets the inner cut-off 
radius, $r_{\rm 0}$, since this 25$\%$ of the total energy or, equivalently, 
$1/4\langle L \rangle$ will be lost for bubble creation and injection within 
this radius. We find that $r_{0}$ roughly turns out to be around $\simeq$ 
20 $\hbox{--}$ 45 kpc depending on the cluster mass $M_{\rm cl}
(\equiv M_{\rm vir})$ and temporal profiles of pressure and density. 
Higher mass clusters have greater $r_{0}$.

The volume cooling rate is calculated using a fit (Nath 2003) to the normalized 
cooling function $\Lambda_{\rm \scriptscriptstyle N}(T)$ for a metallicity 
of $Z/Z_{\rm \scriptscriptstyle \odot}$ = 0.3, as calculated by Sutherland
\& Dopita (1993). This cooling function incorporates the effects of
free-free emission and line cooling. Thus, the volume cooling rate is $\Gamma = n_{\rm
\scriptscriptstyle e}^2 \Lambda_{\rm \scriptscriptstyle N}(T)$, where
$n_{\rm \scriptscriptstyle e}$ = $0.875 (\rho/m_{\rm \scriptscriptstyle
p})$ is the electron density.

\subsubsection{Thermal Conduction}

A potential difficulty in raising the entropy of the intracluster medium 
at large radii ($r\,>\,0.1r_{\rss 200}$) by means of a central heating source 
is that the energy required is fairly large ($\ga\,10^{62}$ erg) over a period 
of the age of the cluster. This sets up a negative entropy 
gradient in the central regions of clusters (see Figure~2 in RRNB04) and a 
rising temperature profile (see right panel of Figure~3 in RRNB04) which are 
somewhat different from
the observed temperature and entropy profiles. 
Our previous analysis, however, did not include the effect of thermal
conduction which would have decreased the temperature gradient in the inner
region and made it consistent with observations. We, therefore, include the
effect of thermal conduction here to address these issues.

Thermal conduction has been suggested
 to be an important process in galaxy clusters 
for quite some time (Bertschinger \& Meiksin 1986; Malyshkin 2001; Brighenti 
\& Mathews 2002; Voigt \etal 2002; Fabian, Voigt \& Morris 2002). However it  
is not clear as to how dominant it would be since magetic fields would 
suppress the conduction co-efficient by a large amount from the classical Spitzer 
value. However, recent theoretical works by Narayan \& Medvedev (2001), Chandran 
\& Maron (2004), Loeb (2002) and several others suggests that conduction
 could be as high as 10\% to 20\% of the Spitzer value in the presence 
of a tangled and turbulent magnetic field. Motivated by these results we adopt 
the suppression factor $f=0.1$.
  
The flux due to thermal conduction $F_{\rm cond}$ is given by

\begin{equation}
\mathbf{F}_{\rm cond}=-f\kappa \nabla T,
\end{equation}

\noindent
where $\kappa$ is the Spitzer conductivity

\begin{equation}
\kappa =\frac{1.84\times 10^{-5}T^{5/2}}{\ln\lambda},
\end{equation}

\noindent
with the Coulomb logarithm $\ln\lambda = 37$, approprite for ICM temperature
and density.
 
\subsection{Evolution of the ICM}

The intracluster gas is assumed to be in quasi-hydrostatic equilibrium at 
all times since cooling is not precipitous at these radii and the heating is mild. 
The gas entropy per particle is
\be
S = {\rm const} + \frac{1}{\gamma -1}k_{\rm
\scriptscriptstyle b}\ln(\sigma).
\ee

\noindent
where $\sigma \equiv P_{\rm \scriptscriptstyle gas}/\rho_{\rm
\scriptscriptstyle gas}^{\rm \scriptscriptstyle \gamma}$ is the ``entropy
index'' and $\gamma$ is the 
adiabatic index. The particle number density of the gas, $n$, is given by
$n = \rho_{\rm \scriptscriptstyle gas}/\mu m_{\rm 
\scriptscriptstyle p}$.

During each timestep $\Delta t$, the entropy of a given mass shell changes by an amount

\be
\Delta S = \frac{1}{\gamma -1} k_{\rm \scriptscriptstyle b} {\Delta \sigma \over
\sigma} = {1 \over {n T}}({\cal H} - \Gamma -
\nabla\cdot F_{\rm \scriptscriptstyle cond})\Delta t .
\label{eq:del_S}
\ee

\noindent
After evaluating the change in the entropy index of each mass shell of gas due to 
heating, cooling and conduction after a time $\Delta t$ using equation~(\ref{eq:del_S}), 
the new entropy index of each shell is calculated using
\be
\sigma_{\rm \scriptscriptstyle new}(M) = \sigma_{\rm \scriptscriptstyle 0}
(M) + \Delta \sigma (M)
\label{eq:sig_update}
\ee

\noindent
where $ \sigma_{\rm \scriptscriptstyle 0}(M) = P_{\rm \scriptscriptstyle
gas0}/\rho_{\rm \scriptscriptstyle gas0}^{\rm \scriptscriptstyle \gamma}$
is the default entropy index. The system relaxes to a new state of
hydrostatic equilibrium with a new density and temperature profile. After
updating the function $\sigma (M)$ for each mass shell, we solve the
equations 
\bea
{dP_{\rm \scriptscriptstyle gas} \over dM}&=& {G M(\le r) \over
{4\pi r^{\rm \scriptscriptstyle 4}}}\label{eq:hydro_b}
\\
\, {dr \over dM} &=& {1 \over 4\pi r^{\rm \scriptscriptstyle 2}}
{\left ({P_{\rm \scriptscriptstyle gas}(M) \over \sigma(M) }\right )}^{(\rm
\scriptscriptstyle 1/\gamma)}
\label{eq:evol_gas}
\eea
to determine the new density and temperature profiles at time $t+\Delta t$.
The boundary conditions imposed on these equations are that (1) the
pressure at the boundary of the cluster, $r_{\rm \scriptscriptstyle out}$,
is constant and is equal to the initial pressure at $r_{\rm
\scriptscriptstyle 200}$, i.e., $P\,(r_{\rm \scriptscriptstyle out})$ =
constant = $P_{\rm \scriptscriptstyle gas0}$($r_{\rm \scriptscriptstyle
200}$), and (2) the gas mass within $r_{\rm \scriptscriptstyle out}$ at all
times is the mass contained within $r_{\rm \scriptscriptstyle 200}$ for the
default profile at the initial time, i.e., $M_{\rm \scriptscriptstyle g}
(r_{\rm \scriptscriptstyle out}) = M_{\rm \scriptscriptstyle g0}(r_{\rm
\scriptscriptstyle 200}) = 0.1333M_{\rm \scriptscriptstyle dm} (r_{\rm
\scriptscriptstyle 200})$. It is important to note here that 
$r_{\rm \scriptscriptstyle out}$ increases as the cluster gas gets heated and 
spreads out.

The observed gas entropy ${\cal S}$($r$) at $0.1r_{\rm \scriptscriptstyle
200}$ and at $r_{\rm \scriptscriptstyle 500}$ is then calculated using 

\be
{\cal S}(r)\equiv T(r)/n_{\rm \scriptscriptstyle e}^{2/3}(r) .
\ee

\noindent
The updated values of $\sigma (M)$ and pressure of the ICM $P_{\rm
\scriptscriptstyle gas}(r)$ are used to calculate the heating and cooling
rates and the conduction flux for the next time step. This is continued for
a duration of $t_{\rm \scriptscriptstyle heat}$.  After that the heating
source is switched off, putting ${\cal H}$ = 0. 
The cooling rate and conduction flux continue to be calculated to update
the function $\sigma (M)$ at subsequent timesteps, and the hydrostatic
structure is correspondingly evolved for a duration of $t_{\rm
\scriptscriptstyle H} - t_{\rm \scriptscriptstyle heat}$, where $t_{\rm
\scriptscriptstyle H}$ is the Hubble time. Note that $r_{\rm
\scriptscriptstyle out}$ decreases during this time since the intracluster
gas loses entropy and shrinks. 
The only free parameters in our calculation are the energy injection rate 
\lLr and the time $t_{\rm \scriptscriptstyle heat}$ over which the
``effervescent heating'' of the ICM takes place.
After evolving the gas for the total available time, $t_{\rm
\scriptscriptstyle H} \sim 1.35\times 10^{10}$ years, we check whether the
entropy at $0.1r_{\rm \scriptscriptstyle 200}$ and $r_{\rm
\scriptscriptstyle 500}$ match the observed values, and adjust parameters
accordingly. In this way, we explore the parameter space of \lLr and
$t_{\rm \scriptscriptstyle heat}$ or rather a single free parameter, i.e. the 
total energy, $E_{\rm agn}=\langle L\rangle t_{\rss heat}$ for different cluster masses so 
that the entropy (after 1.35$\times$10$^{10}$ years) at $0.1r_{\rm
\scriptscriptstyle 200}$ and $r_{\rm \scriptscriptstyle 500}$ matches the
observed values. 

For numerical stability of the code, the conduction term is integrated
using timesteps that satisfy the appropriate Courant condition. The Courant
condition for conduction is 
\be
\Delta t_{\rm \scriptscriptstyle cond} \le 0.5 {(\Delta r)^{2} n\, k_{\rss b} \over 
{\kappa (\gamma-1)}} .
\ee

\noindent
The timesteps, $\Delta t$,  used in equation~(12) to update the entropy of
the gas and calculate its pressure, temperature and density 
profiles always obey the above Courant condition (Ruszkowski \& Begelman
2002; Stone, Pringle \& Begelman 1999). 

\section{Thermal Sunyaev-Zeldovich effect}  

The temperature decrement of CMB due to the SZ effect is directly proportional 
to the Compton parameter ($y$). For a spherically symmetric cluster, the 
Compton parameter is given by

\begin{equation}
y=2\frac{\sigma_{\rm T}}{m_{\rm e}c^2}\int_{0}^{R}p_{\rm e}(r)dl
\label{eq:y_sph_sym}
\end{equation}

\noindent
where $\sigma_{\rm T}$ is the Thomson cross-section, and $p_{\rm e}(r)$ =
$n_{\rm e}(r)k_{\rm b}T_{\rm e}(r)$ is the electron pressure of the ICM,
where $n_{\rm e}(r)$ = $0.875 (\rho_{\rm gas}/m_{\rm p})$ is the 
electron number density, $k_{\rm b}$ is the Boltzmann constant, and $T_{\rm e}(r)$ is 
the electron temperature. The integral is performed along the line$\hbox{--}$of
$\hbox{--}$sight ($l$) through the cluster and the upper limit of the integral ($+R$) 
is the extent of the cluster along any particular line$\hbox{--}$of$\hbox{--}$sight.
We do not include the effects of beam size in calculating the $y$ parameter. This
approximation is justified by the fact that the pressure profiles are relatively 
flat in the inner region. The variation of pressure integrated along the line--of--sight
as a function of the projected radius is even flatter 
thus providing more justification for the above approximation.

The angular temperature profile projected on the sky due to SZ effect, 
$\Delta T(\theta)/T_{\rm CMB}$ is given in terms of the Compton 
parameter in equation~(\ref{eq:y_sph_sym})

\begin{equation}
\frac{\Delta T(\theta)}{T_{\rm CMB}}=g(x)y(\theta),
\end{equation}

\noindent
where $g(x)\equiv x$coth($x/2$)-$4$, $x\equiv h\nu/k_{\rm B}T_{\rm CMB}$, 
$T_{\rm CMB}=2.728$ (Fixsen et al. 1996). In the Rayleigh-Jeans approximation, $g(x)\approx -2$.  
We only evaluate ``central'' SZ decrement from the pressure profiles of our models.
In this case, the integral in equation~(\ref{eq:y_sph_sym}) reduces to

\begin{equation}
y_{\rm 0}=2\frac{\sigma_{\rm T}}{m_{\rm e}c^2}\int_{0}^{R}p_{\rm e}(r)dr
\label{eq:y0}
\end{equation}

\noindent
In the Rayleigh-Jeans part of the CMB spectrum, the deviation from the black-body spectrum 
results in a decrement of the CMB temperature,

\begin{equation}
\Delta T_{\rm \mu w0} \approx -5.5\,y\,K
\label{eq:del_T}
\end{equation}

\noindent
We use the pressure profiles resulting from our model to calculate the
central SZ decrement in the temperature of the CMB.


\begin{figure*}
\centering
\includegraphics[width=3in]{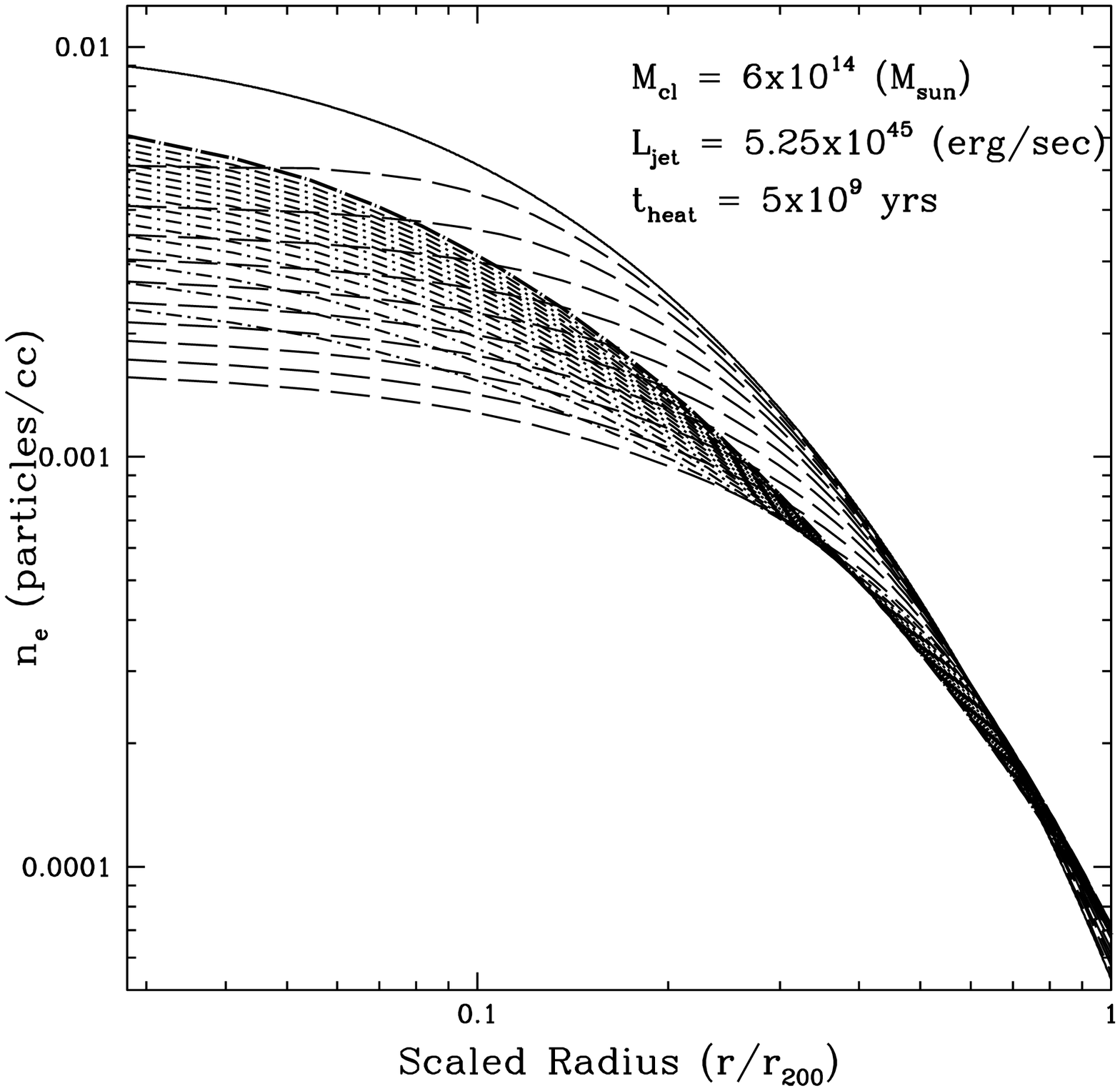}
\includegraphics[width=3in]{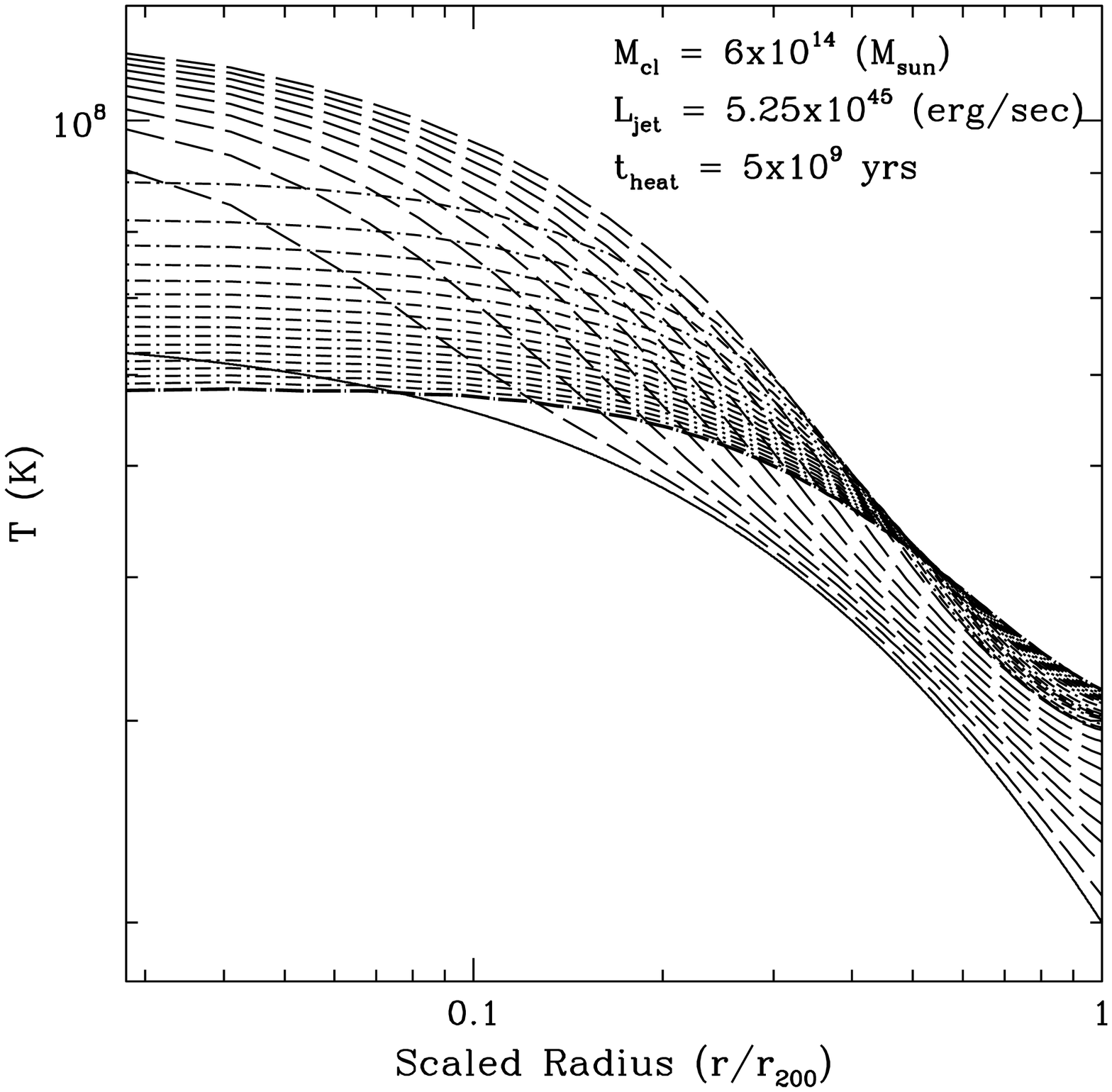}
\caption{\small Gas density (left panel) and temperature (right panel) profiles
as a function of scaled radius ($r/r_{\rm \scriptscriptstyle 200}$), for the 
cluster model described in the text. Dashed lines represent density profiles when 
heating is active and the dot-dashed line represents density profiles after the 
heating source has been switched off (i.e. after $t_{\rss heat}\,>\,5\times10^9$ years). 
The profiles are plotted after every $5\times10^8$ years till Hubble time. It is seen 
here that conduction removes temperature gradients in the central regions (within 
$0.2r_{\rss 200}$) and flattens the temperature profile. This is the reason thermal 
conduction was included in the cluster model as compared to RRNB04, wherein left panel 
of Figure~(3) shows rising temperatures in the centre. The density profiles are seen to rise 
after the heating source is switched off and thermal conduction and radiative 
cooling are the only two processes which are active. Initial density and temperature
profiles correspond to the solid curves in both plots.}
\label{fig:den_temp_cond}
\end{figure*}

\begin{figure*}
\begin{center}
\includegraphics[width=3.5in]{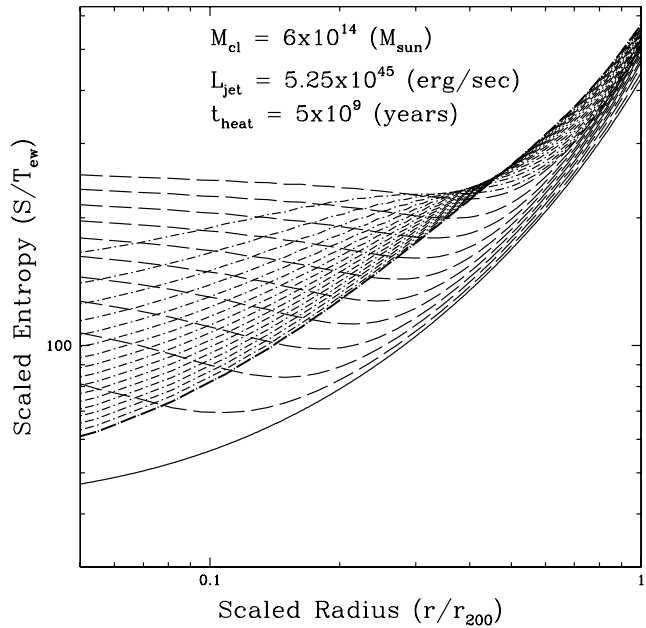}\hfill
\caption{Scaled entropy profiles as a function of scaled radius for a
cluster of mass $6 \times 10^{14}$ $M_{\rm \scriptscriptstyle \odot}$
heated by an AGN with \lLr$=3\times 10^{45}$ erg s$^{-1}$. The scaled entropy
profiles are plotted at intervals of 5$\times$10$^8$ years. The dashed lines 
correspond to times when heating is on and the dot-dashed lines correspond 
to times when heating has been switched off. They are seen to rise as the gas is 
heated and then fall as the gas cools. It is seen here that the final entropy 
profiles (after hubble time) has neither negative entropy gradient nor any entropy 
core in the central regions, as compared to Figure~(2) in RRNB04. Initial states 
correspond to the solid curves.}
\label{fig:ent_cond}
\end{center}
\end{figure*}

\begin{figure*}
\begin{center}
\includegraphics[width=3.5in]{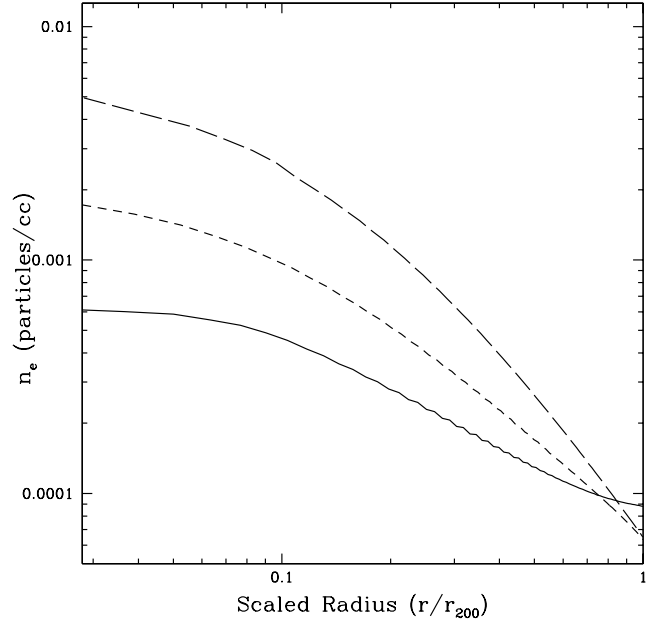}\hfill
\caption{Final density profiles for objects of mass $4.5\times 10^{13}$ M$_{\odot}$ 
(solid line), $2.0\times 10^{14}$ M$_{\odot}$ (short dashed line), 
$9.0\times 10^{14}$ M$_{\odot}$ (long dashed line).       
The assumed heating time $t_{\rm heat}$ in this example was $5\times 10^{9}$ years.
Note the flattening in the 
density profile as the mass of the system is decreased.}
\label{fig:den_cond}
\end{center}
\end{figure*}


\section{Angular power spectrum}

The angular two-point correlation function of the SZ temperature distribution  
in the sky is conventionally expanded into the Legendre polynomials:

\begin{equation}
\Big \langle \frac{\Delta T}{T_{\rm CMB}}({\mathbf {\hat n}})
\frac{\Delta T}{ T_{\rm CMB}}({\mathbf {\hat n}} + {\mathbf {\theta}}) \Big \rangle = 
\frac{1}{4\pi}\sum_{l}(2l+1)C_{l}P_{l}(\cos\theta)
\end{equation}

\noindent
Since we consider discrete sources, we can write $C_{l}= 
C_{l}^{(P)} + C_{l}^{(C)}$, where $C_{l}^{(P)}$ is the contribution 
from the Poisson noise and $C_{l}^{(C)}$ is the correlation among 
clusters (Peebles 1980, \S~41). We define the frequency independent part in the power 
spectrum as $C_{l}^{*(P)}\,\equiv\,C_{l}/g^{2}(x)$. The integral expression
of $C_{l}^{*(P)}$ can be derived following Cole \& Kaiser (1988) as

\begin{equation}
C_{l}^{*(P)} = \int_{\rm 0}^{z_{\rm dec}} dz{dV \over {dz}}
\int_{\rm M_{\rm min}}^{M_{\rm max}} dM {dn(M,z) \over {dM}}
|y_{l}(M,z)|^2,
\label{eq:Cl}
\end{equation}

\noindent
where $V(z)$ is the co-moving volume and $y_{l}$ is the angular Fourier transform 
of $y(\theta)$ given by

\begin{equation}
y_{l}=2\pi\int y(\theta) J_{0}[(l+1/2)\theta]\theta d\theta ,
\end{equation}

\noindent
where $J_{0}$ is the Bessel function of the first kind of the integral
order $0$.
In equation~(\ref{eq:Cl}), $z_{\rm dec}$ is the redshift of photon decoupling and 
$dn/dM$ is the mass function of clusters which is computed in the Press-Schechter 
formalism (Press \& Schechter 1974). The mass function has been computed using the 
power spectrum for a $\Lambda$CDM model with normalization of 
$\sigma_{8}=0.9$. We choose $M_{\rm min}=5\times 10^{13}M_{\odot}$ and $M_{\rm max}
=2\times 10^{15}M_{\odot}$ and integrate till redshift of $z=5$ 
instead of $z_{\rm dec}$. This is done because the integral in 
equation~(\ref{eq:Cl}) is found to be insensitive to the upper limit in redshift 
beyond $z=4$, the reason being that the mass function is exponentially 
suppressed beyond that redshift in this mass range.


\begin{figure*}
\centering
\includegraphics[width=3.5in]{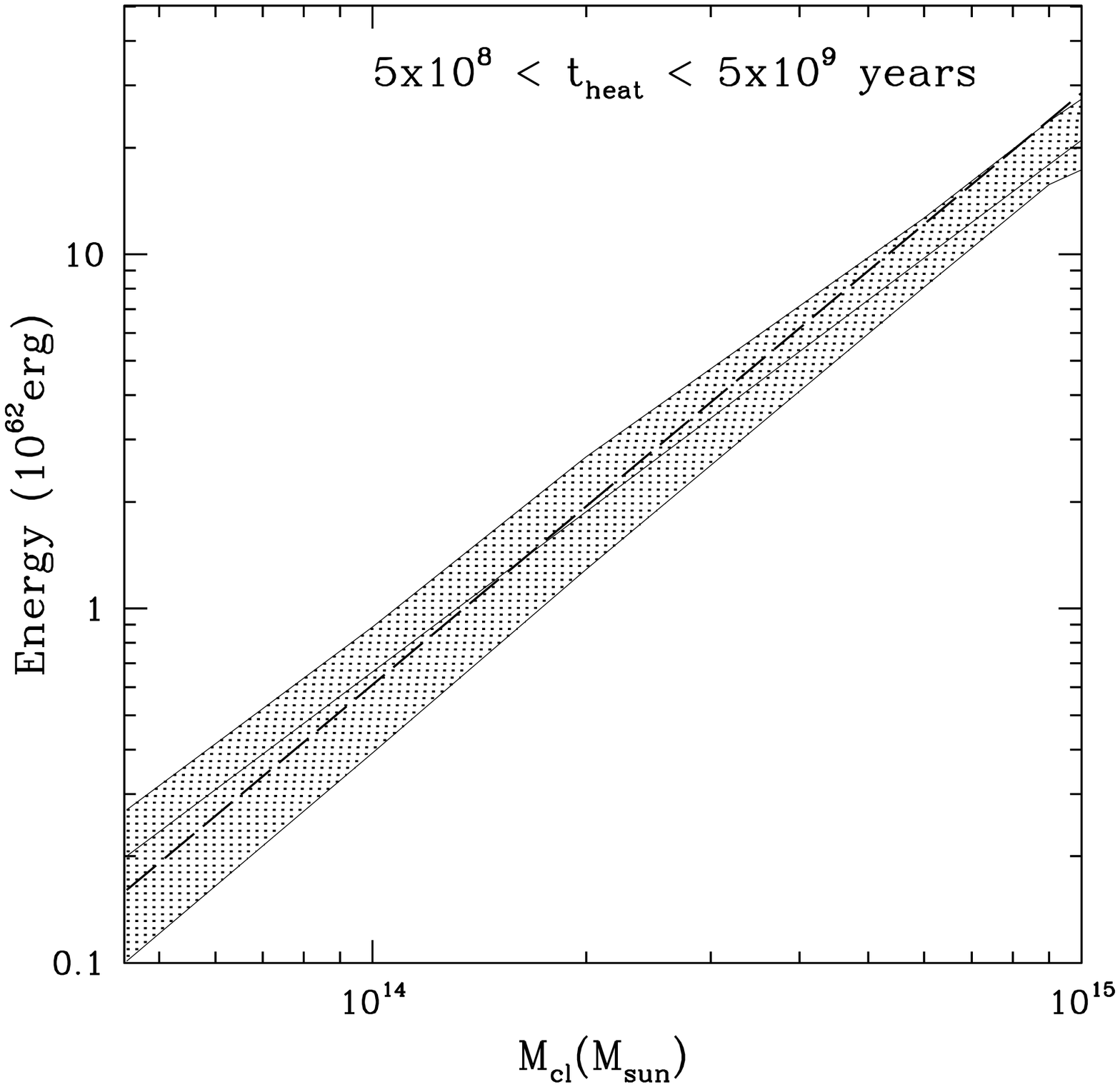}\hfill
\caption{This figure shows the permitted total injected energy range as a
function of the cluster mass for ICM heating times between $t_{\rm
heat}=5\times 10^{8}$yr (upper envelope) and $t_{\rss heat}=5\times 10^{9}$yr
(lower envelope). The shaded region  corresponds to values of $E_{\rm agn}$
that are able to match the entropy observations at {\it both} $0.1r_{\rm
\scriptscriptstyle 200}$  and $r_{\rm \scriptscriptstyle 500}$. The thin
solid line represents a non-linear relation between the total energy injected
into the cluster by AGN and the mass of the cluster with an exponent of 1.5. The 
thick long-dashed line represents a non-linear relation between the total energy 
injected into the cluster by AGN and the mass of the cluster with an exponent of 
5/3. The permitted parameter space comes from the sum of permitted regions 
that satisfy the entropy constraints at both radii for fixed $t_{\rss heat}$.}
\label{fig:E_Mcombined}
\end{figure*}


\section{Results}

In this section, we discuss our results for cluster evolution due to
heating, cooling and conduction. We also discuss our results for the 
central SZ decrement for clusters with masses ranging from $M_{\rm cl}=5\,\times 
10^{13}\hbox{--}2\times 10^{15}M_{\rm \odot}$. 

The gas is heated for a time $t_{\rm\scriptscriptstyle heat}$ and cooled simultaneously. 
After this time, the heating source is switched off. The gas is then allowed to cool 
radiatively until a total simulation time of $t_{\rm \scriptscriptstyle
H}$=1.35$\times$10$^{10}$ years has elapsed. The final entropy values at
$0.1r_{\rm \scriptscriptstyle 200}$ and $r_{\rm \scriptscriptstyle 500}$
are compared with the observed ones. 

In Figure~(\ref{fig:den_temp_cond}), the evolution of the density and temperature 
profiles of the ICM are shown for a cluster of mass $6\times10^{14}$ 
$M_{\rm \scriptscriptstyle\odot}$ and for a luminosity of \lLr $ = 5.25\times 
10^{45}$ ergs$^{-1}$.  The gas density decreases with time during the heating
epoch, and increases due to radiative cooling and conduction after the heating 
source is switched off. It is interesting to note that the changes in density 
are minimal beyond $0.5r_{\rm\scriptscriptstyle 200}$, as compared to $0.2r_{\rss 200}$ 
in Figure~(3) in RRBN04, and that conduction plays a very important role in regulating 
the density profiles after the heating source is switched off. It is seen that conduction 
actually decreases the density of the gas at larger radii (beyond $0.5r_{\rss 200}$) by 
conducting heat out from the central regions. This is seen more clearly if one studies 
the evolution of the temperature profiles. After the heating source is switched off, 
it is seen that the temperature of the central regions fall very rapidly since 
conduction pumps out heat from the central regions and redistributes it in the 
outer regions of the cluster. Thus the temperature profiles do not rise towards 
the centre as compared to what is seen in Figure~(3) in RRBN04. On the other hand, 
their evolution shows a rise in the outer regions (beyond $0.5r_{\rss 200}$) due 
to thermal conduction even after the heat source has been switched off. Thus conduction 
acts like a heating source for larger radii. 

Figure~(\ref{fig:ent_cond}) shows the time evolution of scaled entropy 
profiles of a cluster of mass $M_{\rm \scriptscriptstyle cl} = 6\times10^{14}
M_{\rm \scriptscriptstyle \odot}$ for \lLr = 5.25$\times 10^{45}$ erg s$^{-1}$. 
We use the same method of emissivity weighting as in Roychowdhury \& Nath (2003) 
to calculate the average quantities. The entropy profiles are plotted in 
time-steps of $5\times10^{8}$ years. They are seen to rise with time as the
ICM is heated. Then, after the heating is switched off (after
$t_{\rm \scriptscriptstyle heat} = 5\times 10^9$ years), the gas
loses entropy due to cooling and the profiles are seen to fall
progressively. The inclusion of
conduction removes the negative gradient in the scaled entropy
profiles in the central regions of the cluster (within $0.5r_{\rm
\scriptscriptstyle 200}$. These entropy profiles do not show any 
flat entropy core unlike in RRBN04 (left panel in their Figure 2). Thus, 
this probably indicates that thermal conduction is a more plausible 
process in ICM than covection for such gentle AGN heating.

The model is constrained by entropy-temperature relation.
At fixed cluster temperature this corresponds to a given density 
(at both radii for which the entropy data is provided). 
Now, since at a given density we have an additional 
constraint from the L-T relation (that our model fits reasonably well; see below), 
we implicitely
satisfy the constraints on the slope of the density profile. Thus,
as the model fits both the entropy data and the observed L-T relation that 
specify the slopes of the density profiles, these slopes must also be consistent with 
observations that show flattening in low mass systems. Indeed, this flattening is apparent 
in Figure 3 that shows final density profiles for different masses.
 
We now discuss the permitted range in the total energy injected
into the cluster, $E_{\rm \scriptscriptstyle agn} =$ \lLr
$\times\, t_{\rm \scriptscriptstyle heat}$, required to match the
observed entropy as a function of the cluster mass. 

Figure~(\ref{fig:E_Mcombined}) shows the permitted total injected energy 
range as a function of the mass of cluster for heating times between 
$t_{\rss heat}=5\times 10^{8}$ years and $t_{\rss heat}=5\times 10^{9}$ years. Here 
the entropy is required to match observations at {\it both} $0.1r_{\rm 
\scriptscriptstyle 200}$ and $r_{\rm \scriptscriptstyle 500}$. The thick solid 
line represents a relation between the total energy injected to the 
cluster by AGN and the mass of the cluster (see next section for more details).


\begin{figure*}
\begin{center}
\includegraphics[width=3.5in,angle=-90]{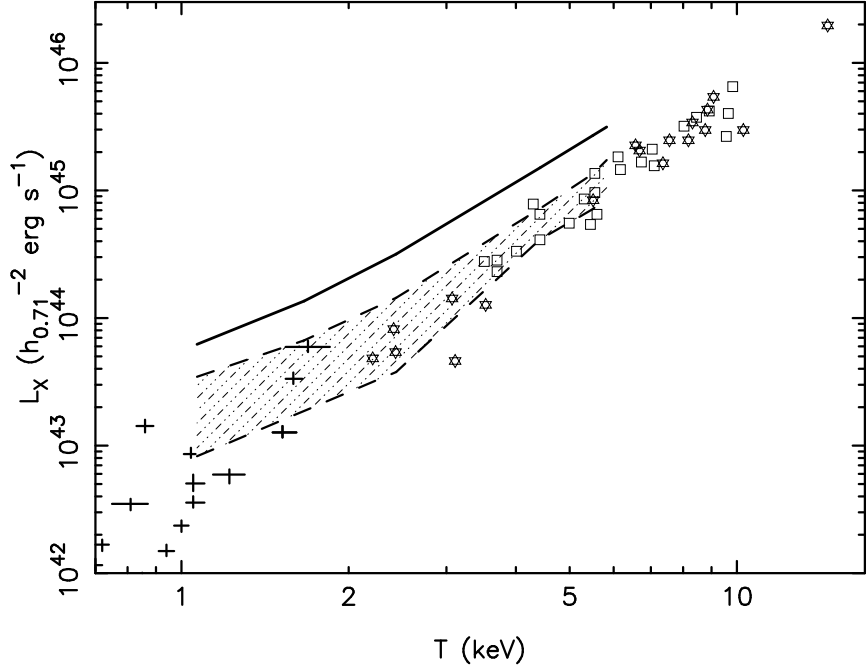}\hfill
\caption{Relation between bolometric X-ray luminosity $L_X$ and emission-
weighted temperature ($\langle T \rangle$). The data points represented
by 'stars' show measurements of clusters with insignificant cooling flows
compiled by Arnaud \& Evrard (1999). Open squares show cooling
flow-corrected measurements by Markevitch \etal (1998). The data points
with error bars show group data from Helsdon \& Ponman (2000). The shaded
region represents X-ray luminosity calculated for the region of $E_{\rss agn}$ 
shown in Figure~(\ref{fig:E_Mcombined}) which satisfies the observed entropy 
requirements at both radii for $5\times10^8\,<\,t_{\rss heat}\,<\,5\times10^9$ 
years. The solid line represents the X-ray luminosity calculated using 
the default model of the ICM ie. the universal temperature profile. 
The models assume a $\Lambda$CDM cosmology with $\Omega_M$ = 0.29, 
$\Omega_{\Lambda}$ = 0.71, and $\Omega_b$ = 0.047, and a Hubble parameter of 
$h$ = 0.71 has been applied to the models and the data.
}
\label{fig:lum}
\end{center}
\end{figure*}


Figure~(\ref{fig:lum}) shows the X-ray luminosity ($L_{\rss X}$ versus 
emission-weighted temperature ($T_{\rss X}$)relation in clusters. The data points 
have been compiled from Arnaud \& Evrard (1999) (represented by stars), Markevitch 
\etal 1998 (represented by open squares) and Helsdon \& Ponman 2000 (with error bars). 
The X-ray luminosity has been calculated within the cluster volume of $0.3r_{\rss 200}$, 
as done for the data. It is also seen that the X-ray luminosity does not change 
much (within 1\%) if the volume increased from $0.3r_{\rss 200}$ to $r_{\rss 200}$. 
The shaded region in the plot corresponds to the predicted X-ray luminosity when the 
cluster is heated by an AGN for $5\times 10^8\,\le t_{\rm heat}\le \,5\times 10^9$ 
years with 
luminosities which correspond to the two bouding lines of the shaded region in 
Figure~(3). It is seen that the predicted luminosities of the heating model satisfy 
the data points in the low mass end as well as the high mass end. The solid line in the 
plot shows the predicted luminosity due to the universal temperature profile and the 
default density profile. We note that the X-ray luminosity is over-predicted by the 
universal temperature profile which indicates that the addition of non-gravitational 
heating is required to lower the X-ray luminosity to satisfy the data points
(see Roychowdhury \& Nath 2003, for more details).

\begin{figure*}
\begin{center}
\includegraphics[width=3.5in,angle=-90]{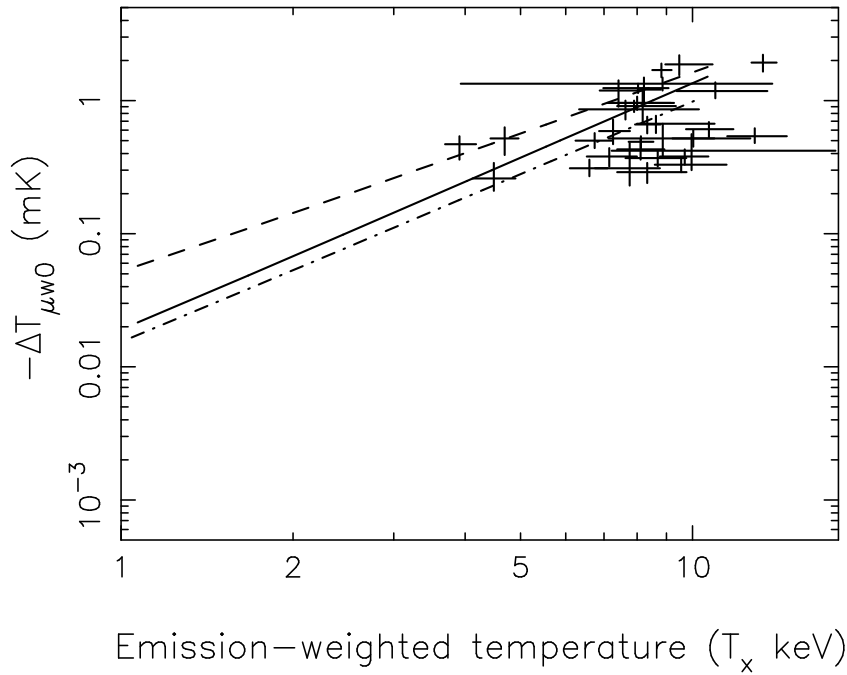}\hfill
\caption{Observed and predicted $\Delta T_{\rm \mu w 0}\,\hbox{--}\langle T\rangle$ 
relation 
of clusters. The solid line represents the predicted $\Delta T_{\rm \mu w 0}$ with 
``universal'' temperature profile (Loken et al. 2002), dark matter density 
profile given by NFW with $r_{\rm c}=0$ and the resulting density profile 
(Roychowdhury \& Nath 2003). The dash-dotted line is the result of the dark matter 
profile given by NFW with $r_{\rm c}=r_{\rm s}/20$ and the ICM temperature profile 
as before. The dashed line is the result of self-similar profile (Wu \& Xue 2002b). 
The data points are from Zhang \& Wu (2000) and McCarthy et al. (2003b).}
\label{fig:y0_def}
\end{center}
\end{figure*}

In Figure~(\ref{fig:y0_def}), the central SZ temperature decrement $\Delta 
T_{\rm \mu w 0}$ is plotted as a function of the emission-weighted temperature 
of the cluster $\langle T\rangle$. The data points are a compilation of data sets from 
Zhang \& Wu (2000) and McCarthy et al. (2003b). The solid line shows the predicted 
$\Delta T_{\rm \mu w 0}$ from the default temperature profile and NFW potential. 
The dot-dashed line shows the predicted $\Delta T_{\rm\mu w0}$ for the same 
temperature profile but for smoothed NFW potential with $r_{\rm c}=r_{\rm s}/20$. 
The dashed line shows the prediction from the self-similar profile 
(Wu \& Xue, 2002b; Bryan 2000).  


It has been shown by Roychowdhury \& Nath (2003) that the density profile of gas 
is much flatter in comparison to the self-similar profile when it assumes the 
``universal temperature profile'' and the standard NFW profile is assumed. 
As a result, the predicted central temperature decrement 
($\Delta T_{\rm \mu w 0}$) is lower than that predicted by the self-similar 
model. The normalization of $\Delta T_{\rm \mu w 0}$ for
a smoothed NFW profile with a core radius $r_{\rm c}\,=
\,r_{\rm s}/20$ is even lower. 
This happens because the introduction of a core radius 
in the dark matter profile makes the ICM density profile shallower in the central 
regions as compared to the ICM density with a standard NFW profile. These decrements 
are closer to the data than predicted by earlier self-similar models for rich clusters. 
 
Next we examine the effects of the effervescent heating, radiative 
cooling and conduction on the central SZ decrement. We 
evaluate $\Delta T_{\rm \mu w 0}$ for clusters in our sample after they 
have been evolved for a Hubble time $t_{\rm H}$. 
The heating source was active for $t_{\rss heat}\ll\,t_{\rm H}$. 
The values of $\langle L\rangle$ and $t_{\rss heat}$ have been chosen 
so as to satisfy observational constraints on ICM entropy  
after $t_{\rss H}$ at $0.1\,r_{\rss 200}$ and 
$r_{\rss 500}$ (Ponman et al. 2003; see Figure 4 and 5 in RRNB04). In other words, 
there is a range of $\langle L\rangle$ that satisfies the 
entropy observations at a 1-$\sigma$ uncertainty level that we used 
in our calculations. We used 
smoothed NFW profile with the core radius $r_{\rm c}=r_{\rm s}/20$. 


\begin{figure*}
\begin{center}
\includegraphics[width=3.5in,angle=-90]{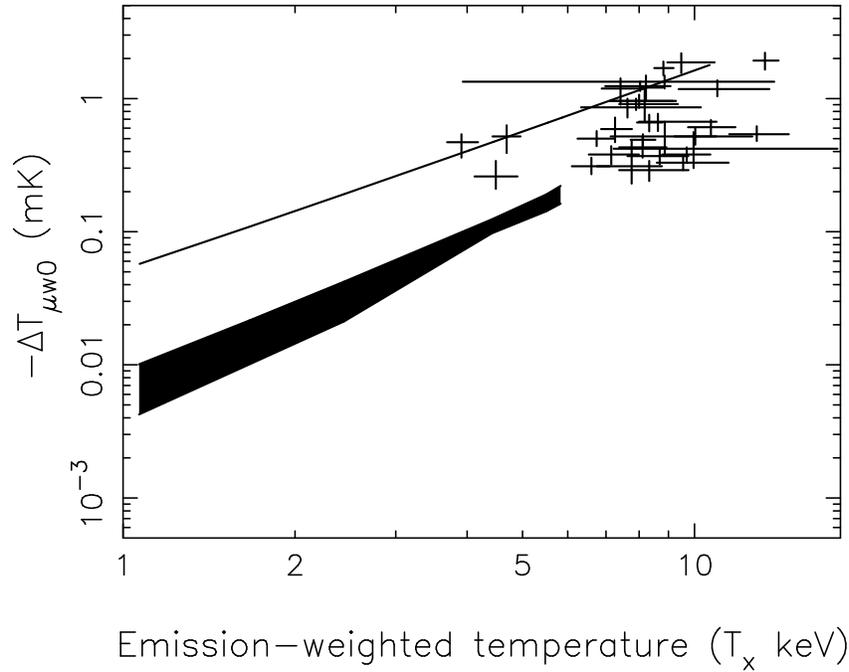}\hfill
\caption{Observed and predicted $\Delta T_{\rm \mu w 0}\hbox{--}\langle T\rangle$ relation 
of clusters. The solid line shows $\Delta T_{\rm \mu w 0}$ for self-similar profile 
(Wu \& Xue 2002b). The shaded regions are the SZ decrement due to heating of the 
ICM, the region shaded with dashed line representing the spread in $\langle L\rangle$ 
which satisfy 
the entropy observations at both radii, $0.1r_{\rss 200}$ and $r_{\rss 500}$ for 
$t_{\rss heat}=5\times 10^9$ years and the gray region representing 
the spread in $\langle L\rangle$ for $t_{\rss heat}=5\times 10^8$ years. 
The data points are from Zhang \& Wu (2000) and McCarthy et al. (2003b).}
\label{fig:y0_heat}
\end{center}
\end{figure*}


In Figure~\ref{fig:y0_heat}, the shaded region is the expectation for the SZ central 
decrement when the gas is heated by the central AGN. The shaded region represents the 
spread in $\langle L\rangle$ which satisfies the entropy requirements at both radii, 
$0.1r_{\rss 200}$ and $r_{\rss 500}$ for $5\times10^8<t_{\rss heat}<5\times 10^9$ years.
The solid line shows the prediction from self-similar profile (Wu \& Xue, 
2002b; Bryan 2000). 

\begin{figure*}
\begin{center}
\includegraphics[width=3.5in,angle=-90]{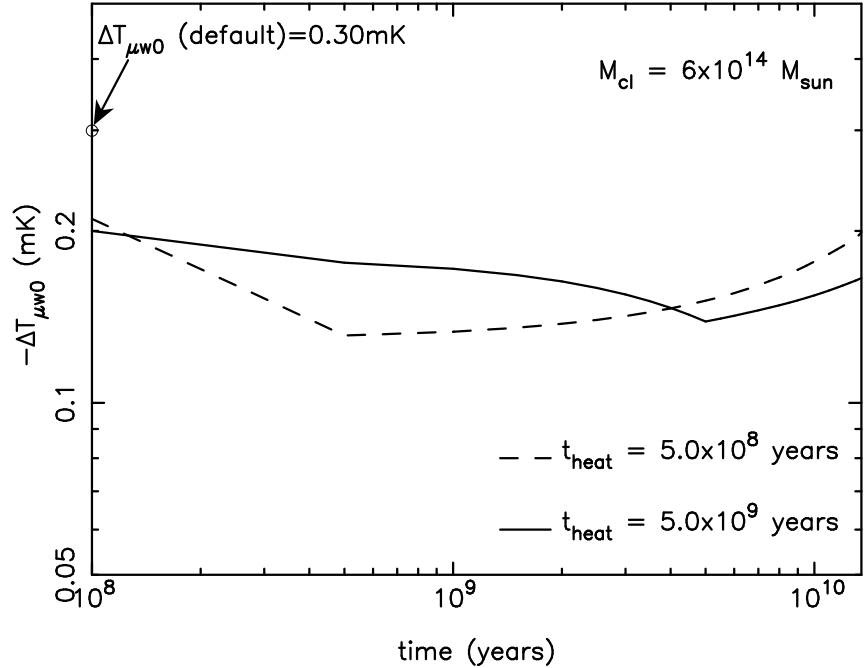}\hfill
\caption{The evolution of the central SZ temperature decrement, $\Delta T_{\rm \mu w 0}$, 
is shown for two different $L_{\rm agn}$ corresponding to the two values 
of $t_{\rss heat}$ 
for $M_{\rm cl}=6\times 10^{14}M_{\rm \odot}$. The point denoted by an open 
circle is the value of $\Delta T_{\rm \mu w 0}$ for the default profile with smoothed NFW 
ie. with a core for the same mass cluster.} 
\label{fig:y0_evol}
\end{center}
\end{figure*}

In Figure~\ref{fig:y0_evol}, the evolution of the central SZ 
decrement $\Delta T_{\mu\rm w0}$ is shown as a result of heating, cooling and conduction 
for a cluster of mass $M_{\rm cl}=6\times 10^{14}M_{\odot}$. 
The dashed line is the result of heating for $t_{\rss heat}=5\times 10^8$ 
years and the solid line is the result of heating the ICM for $t_{\rss heat}
=5\times 10^9$ years. 
It can be seen that, as long as the heating source is active, $\Delta T_{\mu\rm w0}$ 
decreases. When the source is switched off, $\Delta T_{\mu\rm w0}$ starts to increase
as the gas evolves only due to radiative cooling and conduction. 
This happens because the density, or 
equivalently the electron pressure, decreases when the gas is heated but becomes larger 
when it is allowed to cool. In addition, we have also plotted the default $\Delta T_{\rm 
\mu w0}$ for the same cluster with a point denoted with an open circle. 
The central SZ decrement 
corresponding to our heating model is always lower than the default value.


\begin{figure*}
\begin{center}
\includegraphics[width=3.5in,angle=-90]{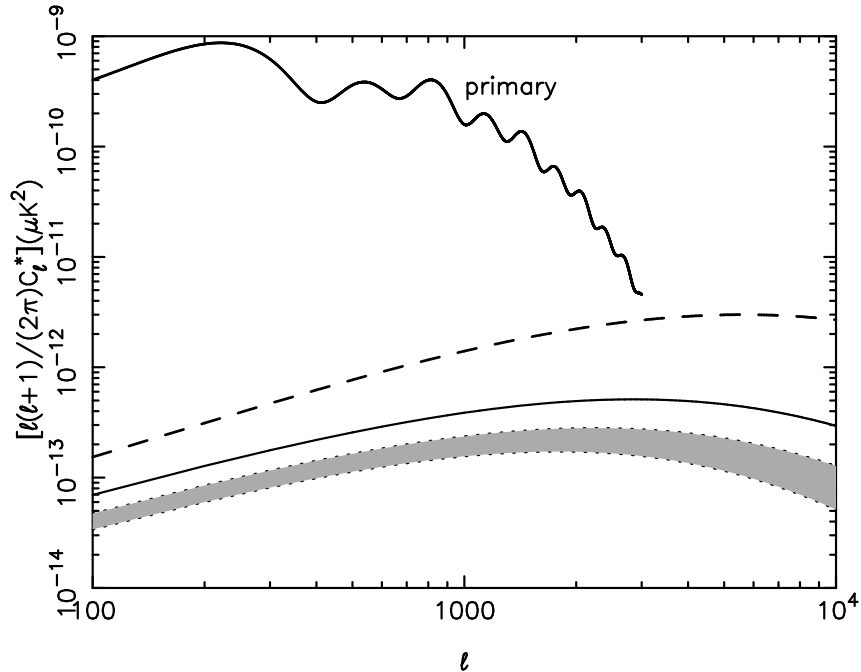}\hfill
\caption{The Poisson contribution to the angular power spectrum ($C_{l}^{*}$) of 
the SZ fluctuations is plotted here as a function of $l$. The thick solid line is for 
the 
primary temperature anisotropy expected in the Rayleigh-Jeans band. The thin solid line 
is for the angular power spectrum (Poisson) due to the ``universal temperature profile'' 
of the cluster ICM and the dashed line is the expectation from a $\beta\hbox{--}$ 
profile and 
isothermal temperature profile. The dotted and the dot-dashed line represent the heated 
ICM models. The dot-dashed line being for $t_{\rss heat}=5\times 10^{8}$ 
years and the dotted line is for $t_{\rss heat}=5\times 10^{9}$ years.}
\label{fig:Cl}
\end{center}
\end{figure*}

Finally, we evaluate the Poisson contribution to the angular power spectrum of the SZ 
and compare our model predictions with earlier ones 
from self-similar models. In Figure~(\ref{fig:Cl}), the thin solid line represents the 
angular power spectrum (Poisson) for the universal temperature profile and the 
corresponding density profile (Roychowdhury \& Nath, 2003). The dashed line is for 
self-similar model (Komatsu \& Kitayama, 1999). The shaded regions represent 
the angular power spectra from heating model. The region shaded with gray represents 
the result due to heating for $t_{\rss heat}=5\times 10^9$ years and the region shaded 
with dashed lines represents the result due to heating for $t_{\rss heat}=5\times 10^8$ 
years. The spread reflects the fact that there is a range of $\langle L\rangle$ for a given
$t_{\rss heat}$ that satisfies the observational entropy data at 
$0.1r_{\rss 200}$ and $r_{\rss 500}$.
  

\section{The halo-black hole mass relation}
The relation between the mass of the group or cluster halo and the total injected enegy
(Figure 4) can be translated to the halo-black hole mass relation. We follow the arguments 
in Wyithe \& Loeb (2003) to derive this relation. 
\indent
Mass of a virialized halo can be 
expressed in terms of its circular velocity $v_{\rm c}$ (Barkana \& Loeb 2001)

\begin{equation}
M_{\rm halo} = 4.7\times 10^{14}\left(\frac{v_{c}}{10^{3}\;{\rm km\; s}^{-1}}\right)^{3} M_{\odot}
\end{equation}

\noindent
We can combine the above relation with the self-regulation condition (Wyithe \& Loeb 2003)

\begin{equation}
\eta L_{\rm Edd}F_{q} = \frac{0.5(\Omega_{b}/\Omega_{m}) M_{\rm halo}v_{c}^{2}}{t_{\rm dyn}},
\end{equation}

\noindent
where $L_{\rm Edd}$ is the Eddington luminosity of the central black hole, 
$\eta$ is its Eddington fraction, $F_{q}$ is the fraction of energy generated by the 
black hole that is deposited in the gas, 
$\Omega_{b}$ and
$\Omega_{m}$ are the mass density in baryons and in mass relative to the critical 
density, respectively. We assumed that the dynamical time of the halo gas is 
$t_{\rm dyn}\sim r_{\rm vir}/v_{c}$, where $r_{\rm vir}$ is the virial radius
of the collapsed halo (Barkana \& Loeb 2001). Combining equations (25) and (26) we get

\begin{equation}
M_{\rm bh}= 2.2\times 10^{8}\left(\frac{\eta F_{q}}{0.05}\right)^{-1}M_{14}^{5/3}M_{\odot},
\end{equation}

\noindent
where $M_{14} = M_{\rm halo}/10^{14}M_{\odot}$. Note that the above relation between
the mass of the cluster and the central black hole is nonlinear. It is interesting that a 
similar scaling relation would be expected in the case of the galaxy halo mass 
$M_{\rm gh}$ and black hole mass relation.
In such a case, the mass of the black hole scales like $M_{\rm bh}\propto v_{c}^{\alpha}$, 
where $4\la\alpha\la 5$. As $v_{c}^{2}\propto M_{\rm gh}/r_{\rm gh}$ 
and $r_{\rm gh}\propto M_{gh}$, we have $M_{\rm bh}\propto M_{\rm gh}^{\alpha/3}$.
In fact, based of the $M-\sigma$ relation and cosmological simulations, 
Ferrarese \& Ford (2005) derive a relation between the mass of the black hole and
the mass of the galactic halo of the form $M_{\rm}\propto M_{\rm gh}^{1.65}$. 
The slope of this relation is very similar to the one considered here.

\indent
Assuming that that a fraction $\epsilon$
of black hole mass is converted to energy, we can also get the relation between
the injected energy $E_{\rm agn}$ and the cluster mass

\begin{equation}
E_{\rm agn} = 6\times 10^{61}\left(\frac{\epsilon}{0.15}\right)
                             \left(\frac{\eta F_{q}}{0.05}\right)^{-1}
                              M_{14}^{5/3} .
\end{equation}
This relation is the same as the one denoted by thick dashed line in Figure 4.
We stress that the scaling of black hole mass and the mass of the cluster 
derived here is not unique. As Figure 4 demonstrates, relations that have slightly 
different slope are also acceptable. For example a somewhat 
shallower slope of 1.5 is also consistent with the entropy constraints. Nevertheless, 
we emphasize that successful fit requires the relation to be nonlinear with index
steeper than unity.

\section{Discussion}

In this paper, we have examined the effects of effervescent heating by AGN 
in clusters with thermal conduction and cooling in the context of the 
excess entropy requirements at large radii. We have also examined the 
consequences of this heating, cooling and conduction on SZ temperature 
decrement. 
 
As is clear from Figure~(\ref{fig:E_Mcombined}), it is possible to 
heat the ICM with a {\it single} central AGN to match the entropy 
requirements at {\it both} $0.1r_{\rm \scriptscriptstyle 200}$ and $r_{\rm
\scriptscriptstyle 500}$. However, in order to match the entropy
at both radii, the total injected energy $E_{\rm\scriptscriptstyle agn}$, 
for a given value of 
$t_{\rm \scriptscriptstyle heat}$ $\ll$ $t_{\rm \scriptscriptstyle H}$, 
must be tightly constrained.  In fact, our calculations have shown that for 
any value of $t_{\rm\scriptscriptstyle heat} < t_{\rm \scriptscriptstyle H}$, i.e.,
for any heating time (or AGN lifetime), it is always possible to satisfy 
the entropy observations at {\it both} radii with a {\it single} value of the
luminosity \lLr. This is different from the cooling flow problem,
where \lLr must be finely tuned to match the cooling rate
(Ruszkowski \& Begelman 2002), because cooling effects on large
scales are rather mild and, thus, the results depend mostly on the
total injected energy $E_{\rm \scriptscriptstyle agn}=\langle
L\rangle\,\times\,t_{\rss heat}$.  Thus, if we can fit the observed
entropy values for just one pair \lLr and $t_{\rss heat}$, we can
do so for a wide range of such pairs. 

We note here that the inclusion of thermal conduction brings down the energy 
which has to be provided by the AGN over its life-time to satisfy the entropy 
observations at both radii for all heating times as compared to our model in RRNB04. 
In addition, for shorter heating times, the $E_{\rss agn}$ is less or comaparable to 
the energy pumped in for longer heating times. This is in contradiction to our 
findings in RRNB04. This happens here because thermal conduction acts as a heating 
source after the AGN is switched off (for $t_{\rss H}\,-\,t_{\rss heat}$) and raises 
the entropy at large radii (at $r=r_{\rss 500}$). The results are mostly 
sensitive to the {\it total} energy input from the black hole, rather than to \lLr and
$t_{\rss heat}$ separately.  As a consequence, satisfactory fits can be obtained as long 
as the total injected energy falls within a relatively narrow range of values, which 
depends on the cluster mass (Figure~(3)).

Finally, we note that cooling and thermal conduction play important roles in
controlling the heating mechanism so that the entropy profiles broadly
match the observed entropy profiles in clusters (Ponman \etal
2003). Notably, in the later stages of evolution of the gas, after the
heating source is switched off, conduction removes negative entropy gradients 
in the central regions of the cluster. Moreover, there is no entropy core seen in 
the final stages of the evolution of the ICM. Instead we see positive 
entropy gradients as observed in the entropy profiles of galaxy groups 
(Mushotzky \etal 2003, Ponman et al. 2003). 
Unlike previously proposed models, our model predicts that isentropic
cores are not an inevitable consequence of preheating.
However, the clusters that show isentropic core have also been observed
(Ponman et al. 2003). We note that our entropy profiles show a core while 
the source of heating is
active. It is conceivable that the clusters which show entropy
cores are being observed during the active phase of the AGN duty cycle.

As seen in Figure~(\ref{fig:y0_def}), the central SZ decrement for individual 
clusters is diminished if the ``universal temperature profile'' corresponding to  
the gravitational interactions of the ICM with the background dark 
matter is used. This reduces the discrepancy between the predicted 
central SZ signal and the observations compared to the predictions based on the standard
self-similar profiles. The solid line in Figure~(\ref{fig:y0_def}) matches 
the observed points for rich clusters ($\langle T\rangle$ $\le$ 5 keV) 
better than the dashed line, 
indicating that the requirement of non-gravitational heating for rich clusters is lower than 
previously thought. It is important to note here that the introduction of a core radius 
in the dark matter profile brings the SZ temperature decrements down further to match the
observed points. This happens because the pressure profile of the intracluster medium 
becomes shallower in the central regions than for the NFW dark matter profile without 
smoothing.

We also find that the inclusion of effervescent heating, cooling and conduction 
have a significant effect on the SZ signal. The gas in the 
central region is depleted as a result of effervescent heating and thus we see a 
diminution in the SZ signal as a result of AGN heating.
Similar conclusion was reached by McCarthy 
et al (2003a, 2003b) and Cavaliere \& Menci (2001). However, 
in our case, 
there is a spread in the SZ signal. The spread reflects the fact that 
there is a range of energies required to satisfy the observed entropy at $0.1R_{\rm 
200}$ and $R_{\rm 500}$ for a particular $t_{\rss heat}$. 
Shaded regions in figure 2 
are based on fits of the heating model to the observational entropy data.
The region shaded with dashed lines corresponds 
to the predicted SZ signal for the range in $E_{\rm agn}$ 
that satisfies the entropy requirements at both radii for $t_{\rss heat}=5\times
10^9$ years and the region shaded gray corresponds to the 
range in $E_{\rm agn}$ that satisfies the entropy requirements at both radii for 
$t_{\rss heat}=5\times 10^8$ years.
As seen in Figure~\ref{fig:y0_heat} the data for SZ are still not good 
enough to constrain heating models at present. We hope that future SZ observations
of low temperature clusters will constrain the models better in the regime where 
discrepancies between self-similar and heating models are more pronounced.
The capabilities of future experiments like SZA are such that a decrement in the CMB 
temperature of 10 $\mu$K could be detected (S. Majumdar, private communication). 
This would mean that the lowest mass systems that SZA could observe 
would have T$\sim$ 2 keV.
It should be noted here that the parameter $L_{\rm agn}$ and thus $E_{\rm agn}$ estimated 
in this paper are slightly different from those in RRNB04. This is due to 
differences in the effervescent heating mechanism and the dark 
matter profile. The inclusion of a self-consistent method for estimating the inner heating 
cut-off radius, $r_{0}$ and a smoothed NFW profile with a core radius
have changed the range in $E_{\rm agn}$ that satisfies the 
entropy requirements at two different radii, $0.1r_{\rss 200}$ and $r_{\rss 500}$, for the 
two values of $t_{\rss heat}$ stated above.
 
We have also studied the time evolution of the SZ temperature decrement due to AGN heating,
cooling and conduction. It is seen in Figure~(\ref{fig:y0_evol}) that 
the SZ signal is diminished as a result of AGN heating at all times in 
comparison to the default SZ signal. This is in contrast to the transient phases
with {\it enhancement} of SZ signals predicted by Lapi et al. (2003)
owing to strong feedback mechanisms that they assumed. Also, in the case of effervescent 
heating, the SZ decrement is lower than the default case even after the gas has evolved 
for a time $t_H-t_{\rss heat}$ after the heating has been switched off. We note here
that the values of $\langle L\rangle$ and $t_{\rss heat}$ have been chosen so as to
satisfy the entropy requirements after evolving for $t_H$.

Finally, we examine the effect of the universal temperature profile and AGN heating 
on the Poisson part of the angular power spectrum. 
We note that the peak of the SZ power spectrum is somewhat sensitive to the amount of heating 
that is added to the cluster gas. The peak of $C_{\ell}$ is at a 
lower $\ell$ for a larger $t_{\rss heat}$ and at a higher $\ell$ 
for a smaller $t_{\rss heat}$. 
Also, the effect due to heating is larger for higher values of $\ell$. 
In our case the suppression 
takes place because the 
gas depletion from the central regions is more efficient in low mass groups than in 
rich clusters. Therefore, the SZ signal is suppressed more efficiently at smaller scales and,
thus, larger $\ell$. We note that the general trend for 
the Poisson contribution to the spectrum 
to be supressed at higher $\ell$ has been noted by others 
(Holder \& Carlstrom, 1999; Kitayama \& Komatsu 
1999; Springel, White \& Hernquist, 2001; Holder \& Carlstrom 2001, Zhang \& Wu 2003). 
This effect leads to a mismatch between the observed and theoretical spectra (Dawson et al. 2001,
Mason et al. 2003) when preheating required to account for the entropy floor is considered.
In our physically motivated effervescent AGN heating model, 
that fits the X-ray observations of entropy of the ICM, the power spactrum 
at small scales is even lower
than previously thought. This suggests that other sources, such as, e.g., ``dead'' radio 
galaxy cocoons at higher redshifts
(Yamada, Sugiyama \& Silk 1999), should significantly contribute to
anisotropies in the cosmic microwave background at large $\ell$. We emphasize that 
our model deals only with global, average effect of heating on clusters and neglects 
small scale fluctuations in the gas distribution that are associated with heating.
      
\section{Conclusion}

The primary aim of this work was to study the implications AGN heating and thermal 
conduction on the global properties of groups and clusters of galaxies including 
the Sunyaev-Zeldovich effect.

\indent
We have demonstrated that the available entropy data, in conjuction with our feedback
model, 
put constraints on the relation between the total energy injected by the AGN and 
the mass of the cluster 
($E_{\rm agn} \propto M_{\rm cluster}^{\alpha}$). The inferred black hole-halo mass scaling 
($M_{bh}\propto M_{\rm cluster}^{\alpha}$, $\alpha\sim 1.5$) is an analog and extension 
of the similar relation between  the black hole mass and the mass of the galaxy halo that 
holds on smaller scales.
 
\indent
The ``effervescent heating'' mechanism heats the gas in the central regions of 
clusters and makes the gas density profile shallower. This reduces the 
electron pressure of the gas and, thus, reduces the SZ temperature 
decrement. This is in accordance 
with the findings of other authors (Cavaliere \& Menci 2001, McCarthy et al. 
2003b, Lapi at al. 2003). Here we have also shown how the SZ decrement would 
evolve as heating, cooling and conduction regulate the physical state of the ICM.
Our heating model is consistent with the available entropy data at $0.1r_{\rss 200}$
and $r_{\rss 500}$.
We give specific predictions of our model for the SZ decrement for low mass
(low temperature) clusters.
Future observations performed with, e.g., Sunyaev-Zeldovich Array (SZA) or 
Combined Array for Research in Millimiter-wave Astronomy (CARMA) will be able 
to test these predictions.

We also point out that the ``universal temperature 
profile'', that takes into account pure gravitational 
interactions, leads to lower
SZ decrement than that calculated assuming that 
the ICM has a self-similar profile and is in better agreement with data for high 
$\langle T\rangle$
clusters. This implies that the discrepancy between observations and models 
without heating is reduced.

Unlike in the case of previously proposed models, we found that isentropic
cores are not an inevitable consequence of preheating. This is consistent 
with observations of groups that do not show large isentropic cores 
(Ponman et al. 2003, Mushotzky 2003).
Clusters that show isentropic core have also been observed
(e.g., Ponman et al. 2003). We note that our entropy profiles show a core while 
the source of heating is 
active which may explain such cases as well.
It is conceivable that the clusters which show entropy
cores are being observed during the active phase of the AGN duty cycle.
This suggests that there may exist an observationally testable
correlation between the presence of isentropic cores and radio emission from
the central galaxy. However, we note that such a correlation may not be strong
as the radio emission may quickly fade away while the mechanical effects of feedback
may still be present.
 
We have also demonstrated that the model reproduces the observed trend for 
the density profiles to flatten in low mass systems.

Finally, we also estimated the angular power spectrum of the CMB due to the SZ 
effect from Poisson distributed clusters. We showed that the average effect of heating 
is to reduce the SZ signal 
and thus the angular power spectrum. The finding that the power spectrum at large 
$l$ is suppressed is consistent with previous results (e.g., Komatsu \& Kitayama 1999, 
Holder \& Carlstrom 2001). However, our results indicate that the contribution to the power 
spectrum that results from AGN heating, that is consistent with entropy measurements, is lower
than previously thought. This suggests that other sources, such as, e.g., dead radio 
galaxy cocoons at higher redshifts
(Yamada, Sugiyama \& Silk 1999), should significantly contribute to small scale anisotropies
of the cosmic microwave background.

\section{Acknowledgments}
We would like to thank Mark Voit for discussions and many insightful comments that 
helped to improve the paper. We would 
also like to thank Mitch Begelman, 
Tetsu Kitayama, Eiichiro Komatsu and Subhabrata Majumdar, Shiv Sethi 
as well as the referee for their 
helpful comments, suggestions and clarifications. 
MR acknowledges the support from NSF grant AST-0307502 and NASA through
{\it Chandra} Fellowship Award Number PF3-40029 issued by the Chandra
X-ray Observatory Center, which is operated by the Smithsonian
Astrophysical Observatory for and on behalf of NASA under contract
NAS8-39073.

\end{document}